\documentclass[aps, prd, 10pt, twocolumn, nofootinbib,floatfix,superscriptaddress]{revtex4-2}
\pdfoutput=1
\usepackage{subfigure}
\usepackage{epsfig}
\usepackage{amsmath}
\usepackage{amsfonts}
\usepackage{amssymb}
\usepackage{amssymb,amsmath,amsfonts}
\usepackage{booktabs} 
\usepackage{siunitx}  
\usepackage{xfrac}
\usepackage{color}
\usepackage{cancel}
\usepackage[utf8]{inputenc}
\usepackage{graphicx}
\usepackage{dcolumn}
\usepackage{bm}
\usepackage{tikz}
\usepackage{ragged2e}
\usepackage{tabularx}

\usepackage{tcolorbox}
\usepackage{hyperref}
\hypersetup{colorlinks, citecolor=red, linkcolor=bluscuro, urlcolor=bluscuro}
\definecolor{rossos}{cmyk}{0,1,1,0.55}
\definecolor{bluscuro}{rgb}{0.15, 0.2, .85}
\definecolor{bluchiaro}{cmyk}{1,.3,0.,0.1}

\makeatletter
\long\def\@makecaption#1#2{%
  \vskip\abovecaptionskip
  \sbox\@tempboxa{#1: #2}%
  \ifdim \wd\@tempboxa > \linewidth
    \begin{center}
      \parbox{\linewidth}{\justifying #1: #2}
    \end{center}
  \else
    \global\@minipagefalse
    \hb@xt@\linewidth{\hskip 0pt #1: #2\hfil}%
  \fi
  \vskip\belowcaptionskip}
\makeatother

\begin{document}

\preprint{APS/123-QED}

\title{Observational implications of Wald-Gauss-Bonnet topological dark energy}

\author{Maria Petronikolou}
\email{petronikoloumaria@mail.ntua.gr}
 \affiliation{\mbox{National Observatory of Athens, Lofos Nymfon, 11852 Athens, 
Greece}}
\affiliation{Department of Physics, National Technical University of Athens, 
Zografou
Campus GR 157 73, Athens, Greece}

\author{Fotios K. Anagnostopoulos}
\email{fotisanagn@uop.gr}
\affiliation{Department of Informatics and Telecommunications, University of Peloponnese, Karaiskaki 70, 22100, Tripoli, Greece}

\author{Stylianos A. Tsilioukas}
\email{tsilioukas@sch.gr}
\affiliation{\mbox{Department of Physics, University of Thessaly, 35100 Lamia, 
Greece}}
\affiliation{\mbox{National Observatory of Athens, Lofos Nymfon, 11852 Athens, 
Greece}}

\author{Spyros~Basilakos}
\email{svasil@academyofathens.gr}
\affiliation{\mbox{National Observatory of Athens, Lofos Nymfon, 11852 Athens, 
Greece}}
\affiliation{Academy of Athens, Research Center for Astronomy and Applied 
Mathematics, Soranou Efesiou 4, 11527, Athens, Greece}
\affiliation{School of Sciences, European University Cyprus, Diogenes Street, 
Engomi, 1516 Nicosia, Cyprus}

\author{Emmanuel N. Saridakis}
\email{msaridak@noa.gr}
\affiliation{\mbox{National Observatory of Athens, Lofos Nymfon, 11852 Athens, 
Greece}}
\affiliation{CAS Key Laboratory for Researches in Galaxies and Cosmology, 
Department of Astronomy, \\
University of Science and Technology of China, Hefei, 
Anhui 230026, P.R. China}
\affiliation{\mbox{Departamento de Matem\'{a}ticas, Universidad Cat\'{o}lica 
del 
Norte, 
Avda.
Angamos 0610, Casilla 1280 Antofagasta, Chile}}

\begin{abstract}
We investigate the observational implications of Wald-Gauss-Bonnet (WGB) 
topological dark energy, a modified cosmological framework derived from the 
gravity-thermodynamics conjecture applied to the Universe’s apparent horizon, 
with the Wald-Gauss-Bonnet entropy replacing the standard Bekenstein-Hawking 
one. Assuming a topological connection between the apparent horizon and interior 
black hole (BH) horizons, we derive modified Friedmann equations where the 
evolution of dark energy depends on BH formation and merger rates, which are 
approximated by the cosmic star formation rate. These equations introduce an 
additional, astrophysics-dependent contribution to the cosmological constant. We 
test two scenarios—one with a vanishing cosmological constant ($\Lambda$=0) and 
another with a modified $\Lambda$-against late-Universe data (SNIa, BAO, Cosmic 
Chronometers) via a Bayesian analysis. Although the WGB framework is consistent 
with observations, information criteria statistically favor the standard 
$\Lambda$CDM model. An analysis of linear perturbations shows that the growth of 
cosmic structures is nearly indistinguishable from that of $\Lambda$CDM, with 
negligible dark energy clustering and minimal deviation in the effective 
Newton's constant. The standard thermal history is also preserved. In 
conclusion, WGB cosmology presents a phenomenologically rich alternative that 
connects dark energy to black hole astrophysics while remaining compatible with 
current cosmological data.
\end{abstract}

\maketitle

\section{Introduction}\label{Introduction} 
The standard cosmological paradigm, known as the $\Lambda$ Cold Dark Matter 
($\Lambda$CDM) model, has been remarkably successful in describing the evolution 
and large-scale structure of the Universe. However, despite its successes, the 
model still faces challenges at both theoretical and observational levels. On 
the theoretical side, the non-renormalizability of general relativity 
\cite{AlvesBatista:2023wqm} and the cosmological constant problem remain 
unresolved. On the observational side, increasing precision in cosmological data 
has revealed persistent tensions between predictions of $\Lambda$CDM and 
measurements of cosmological parameters \cite{Abdalla:2022yfr}.

The most severe of these is the so-called $H_0$ tension, referring to the 
discrepancy between the present value of the Hubble constant inferred from 
early-Universe probes, such as the Planck CMB data combined with BAO 
measurements \cite{Planck:2018vyg}, and its direct determination from 
late-Universe observations, most notably the distance ladder measurements by 
SH0ES \cite{Riess:2021jrx}. While Planck reports $H_{0} = 67.27 \pm 0.60 ; 
\text{km s}^{-1}\text{Mpc}^{-1}$, SH0ES finds $H_{0} = 74.03 \pm 1.42 ; \text{km 
s}^{-1}\text{Mpc}^{-1}$, a discrepancy that now exceeds $5\sigma$ and resists 
explanation in terms of local systematics 
\cite{Verde:2019ivm,DiValentino:2021izs,Riess:2023bfx}. Another observational 
challenge has been the so-called $\sigma_8$ or growth tension, referring to 
differences in the amplitude of matter clustering inferred from the CMB compared 
to weak-lensing and large-scale structure surveys 
\cite{Heymans:2020gsg,eBOSS:2018cab,BOSS:2016wmc}. While recent results from the 
KiDS-Legacy survey and its joint analysis with DES Y3 suggest that this tension 
is significantly reduced \cite{Wright:2025xka,Stolzner:2025htz}, the possibility 
of residual inconsistencies across probes leaves open the question of whether 
new physics may be required. If these tensions are not the result of unaccounted 
systematic effects, they could signal physics beyond the standard cosmological 
model. A wide range of extensions has been proposed, including modified gravity, 
early dark energy, interacting dark energy, running vacuum models, decaying dark 
matter, and string-inspired scenarios, among others 
\cite{DiValentino:2021izs,DiValentino:2020vnx,Hu:2015rva,Sola:2017znb,
Poulin:2018cxd,Pan:2019gop,Anagnostopoulos:2020lec,Vagnozzi:2019ezj,
Capozziello:2020nyq,Krishnan:2021dyb,Tsilioukas:2023tdw,Tsilioukas:2024tjh, 
Paliathanasis:2025dcr, vanderWesthuizen:2025iam}.

In this work, we study the observational implications of a novel framework for 
addressing these challenges: Wald-Gauss-Bonnet (WGB) topological dark energy 
\cite{Tsilioukas:2024seh}. This scenario arises from applying the spacetime 
thermodynamics conjecture to the apparent horizon of the Universe, replacing the 
standard Bekenstein-Hawking entropy with the Wald-Gauss-Bonnet entropy. Assuming 
that the apparent horizon is topologically linked with interior black hole 
horizons, one obtains modified Friedmann equations that depend on black hole 
formation and merging rates, which can be reasonably approximated by the star 
formation rate. Consequently, the dynamics of WGB cosmology depend both on the 
Gauss-Bonnet coupling constant $\tilde{\alpha}$ and on measured astrophysical 
parameters. Effectively, the WGB mechanism introduces an additional contribution 
to the cosmological constant, which can be either positive or negative. This 
framework exhibits intriguing phenomenology. The modified dynamics reproduce the 
standard thermal history of the Universe, while allowing for quintessence - or 
phantom-like behavior of the effective dark energy (DE) sector. Since phantom 
behavior is known to play a key role in alleviating both the $H_{0}$ and 
$\sigma_8$ tensions 
\cite{Abdalla:2022yfr,Heisenberg:2022lob,Heisenberg:2022gqk}, the WGB mechanism 
offers a promising new perspective. 

The manuscript is organized as follows: In Section~\ref{Modified cosmology 
though Wald Gauss Bonnet entropy}, we briefly present the Wald-Gauss-Bonnet 
modified cosmology and develop the analysis of perturbations. In 
Section~\ref{obs-constr}, we present the observational data, the methodology 
employed and the results for the free parameters of the WGB cosmology. Finally, 
in Section~\ref{Conclusion}, we summarize our findings and discuss future 
directions.

\section{\label{Modified cosmology though Wald Gauss Bonnet entropy}Wald Gauss 
Bonnet Cosmology}

In the following, a concise review of Wald Gauss Bonnet (WGB) cosmology 
\cite{Tsilioukas:2024seh}  will be provided. Moreover, we will investigate the 
possibility of WGB scenario to produce the entire DE sector without assuming a 
cosmological constant $\Lambda$ and an extension of \cite{Tsilioukas:2024seh} 
analysis at the perturbation level will be performed. 

When one 
departures from general relativity by adding higher order terms in the 
gravitational Lagrangian, the black hole (BH) entropy for any diffeomorphism 
invariant theory has been calculated by Wald with the use of the Noether charge 
method. In the special case that the Gauss-Bonnet term $\mathcal{G} = R^2 - 4 
R_{\mu\nu} R^{\mu\nu} + R_{\mu\nu\rho\sigma} R^{\mu\nu\rho\sigma}$ (where 
$R_{\mu\nu\rho\sigma}$ is the Riemann tensor and $R$  the Ricci scalar)  is 
added to 
the usual Einstein - Hilbert action in four dimensional spacetime, namely
\begin{equation}\label{EH GB action}
    S = \frac{1}{16 \pi G} \int d^4 x \sqrt{-g} \left( R + \tilde{\alpha}  
\mathcal{G} \right),
\end{equation}
then the corresponding Wald Gauss Bonnet entropy adds a topological term to the 
standard Bekenstein Hawking one i.e.
\begin{equation}\label{WGB entropy}
   S_{\text{WGB}} = \frac{A}{4G} + \frac{2\pi \tilde{\alpha}}{G} \chi(h), 
\end{equation}
were $\tilde{\alpha}$ is the GB coupling constant, $\chi(h)$ is the Euler 
characteristic of the BH horizon $h$, and   \(A = 4\pi r_{h}^{2}\) is the 
area of the   horizon.  

\subsection{Background evolution}

Let us consider a homogeneous and isotropic 
Friedmann-Robertson-Walker (FRW) Universe, with metric
\begin{equation}
\mathrm{d}s^2 = -\mathrm{d}t^2 + a^2(t)  \left( \frac{\mathrm{d}r^2}{1 - k r^2} 
+ r^2 \mathrm{d}\Omega^2 \right),
\end{equation}
where \(a(t)\) is the scale factor, and \(k = 0, +1, -1\) corresponds to flat, 
closed, and open spatial geometry, respectively. According to the 
gravity-thermodynamic conjecture, one can apply the first law of 
thermodynamics on the apparent cosmological horizon of the Universe 
$\tilde{r}_A = \left(H^2 + k a^{-2}\right)^{-1/2}$
\cite{Jacobson:1995ab,Padmanabhan:2003gd,Padmanabhan:2009vy} and result to the 
Friedmann equations \cite{Cai:2005ra,Akbar:2006er,Cai:2006rs}. However, if one 
follows this procedure  for the case of extended entropies, then one typically 
obtains modified Friedmann equations.

Let us apply the gravity-thermodynamic conjecture using the Wald-Gauss-Bonnet 
entropy. Inserting  \eqref{WGB entropy} into the first law of 
thermodynamics $dE = - T \cdot dS$, with the boundary of the system being  
  the   apparent horizon $\tilde{r}_A$, with  temperature $T_{r_A} = 
1/\left(2\pi \tilde{r}_A\right)$, and considering that the Universe is filled 
with matter of energy density $\rho_{m}$ and pressure $p_{m}$, then one 
obtains \cite{Tsilioukas:2024seh}
\begin{align}\label{model 2nd Friedmann}
    \rho_{m} + p_{m} =& - \frac{1}{4 \pi G} (\Dot{H} - k/a^2)\\ \nonumber
    +& \frac{\tilde{\alpha}}{\pi G}\frac{1}{H} \left(H^2 + k/a^2\right)^2 
\dot{\chi}(\mathcal{H}),
\end{align}
and by integration
\begin{align}\label{model 1st Friedmann}
    H^2 =& \frac{8 \pi G}{3} \rho_{m} + \frac{k}{a^2} + \frac{\Lambda}{3} \\ 
\nonumber
    &+ 4\tilde{\alpha} \int_{0}^t \left(H^2 + k/a^2\right)^{2} \dot{\chi}(H) dt.
\end{align}

Equations   \eqref{model 2nd Friedmann} and \eqref{model 1st Friedmann} are the 
two modified Friedmann equations of the model. It is of interest to note that 
the cosmological 
constant $\Lambda$ appears naturally as an integration constant. In what 
follows, we assume flat Universe, i.e. k=0 in accordance with CMB results 
\cite{Planck:2018vyg} . The topological term $\dot{\chi}(H)$ becomes  non 
trivial under the assumption 
that the apparent horizon is topologically linked with the horizons of the 
interior BHs. Specifically, if one assumes that the overall topology of 
causally connected boundaries
remains effectively conserved, then every time a BH horizon is formed, 
two effective puncture disks are created on the apparent horizon and every time 
two BH horizons 
merge into one, two puncture disks close up \cite{Tsilioukas:2024seh}.

Nevertheless, the above consideration may lead to non-trivial implications
during black-hole mergers. As discussed in
\cite{Liko:2007vi,Sarkar:2010xp}, the horizon of a black hole is
topologically equivalent to a two-dimensional sphere $S^2$. Hence,
during the merger of two black holes, the corresponding horizons merge
as well, leading to a change in the Euler characteristic of the horizon
structure. In particular, if one compares the final merged horizon with
the initial disconnected configuration, one obtains
$\delta\chi_h=\chi_f-\chi_{in}=-2$.

In the context of Einstein--Gauss--Bonnet gravity, where the entropy
contains the topological contribution of Eq.~(\ref{WGB entropy}), such
topology-changing processes may induce a non-trivial variation of the
Wald entropy. In particular, for $\tilde{\alpha}>0$, the topological
contribution to the entropy decreases during the merger process, resulting to a violation of the second law, 
suggesting that topology-changing processes may play a physically relevant role
in gravitational thermodynamics and horizon entropy evolution.

Motivated by these considerations, Ref.~\cite{Tsilioukas:2024seh}
proposed that the topology of the cosmological apparent horizon may be
effectively linked to the formation and merger history of interior black
holes. In this phenomenological picture, every time a black-hole horizon
is formed, two puncture disks are effectively created on the apparent
horizon, while black-hole mergers correspond to the annihilation of two
such punctures. Under this working hypothesis, the second law violation can be resolved, the Euler characteristic
of the apparent horizon becomes dynamical and depends effectively on the 
competition between black-hole formation and merger rates.

Within this phenomenological framework, the variation of the apparent-horizon
Euler characteristic can be written as
\begin{equation}\label{delta chi apparent horizon}
    \delta \chi(\mathcal{H})= -2 \;(\delta N_{form} - \delta N_{merg}),
\end{equation}
and thus it depends on the BH formation and merging rate.
We stress that the above relation should be viewed as an effective
phenomenological prescription motivated by gravitational thermodynamics and
topological considerations, whose observational consequences are investigated
in the present work.
Now, the BH formation and merging rate
  can 
be 
approximated by the star formation rate (SFR) best fit model by Madau and 
Dickinson \cite{Madau:2014bja} as
\begin{equation}\label{psi Madau}
   \psi(z) = 0.015 \frac{\left(1 + z\right)^{2.7}}{1 +  \left[\left(1 + 
z\right)/2.9\right]^{5.6}}\;M_\odot \text{year}^{-1} Mpc^{-3},
\end{equation}
where as the dynamical variable we have used the redshift $z$  
defined as $a = (1 + z)^{-1}$, where the scale factor at present is set to 
$a_{0} = 1$ and thus $z_{0} = 0$.
If one assumes that only a fraction of   stars   $f_{BH}$ will become BH 
progenitors  with average mass $\langle m_{\text{prog}} \rangle$, and that from 
the formed BHs only a fraction of them $f_{bin}$  will be in binary systems, 
and that only a fraction of them  will eventually merge $f_{merge}$, then the 
rate of active BHs inside the apparent horizon, defined as $N\equiv N_{form} - 
N_{merge}$ has been calculated in terms of the redshift, for a flat universe $(k 
= 0)$, as  
\cite{Tsilioukas:2024seh}
\begin{align}\label{active dN/dz}
    \frac{dN(z)}{dz} =& \frac{4 \pi}{3} \left( 1 - f_{\text{bin}}  \times 
f_{\text{merge}}\right) \times f_{\text{BH}} \\ \nonumber
    &\frac{ \psi(z)}{\langle m_{\text{prog}} \rangle H^{4}(z) (1 + z)}.
\end{align}
According to the literature, the parameters appearing in
the above expression are estimated within the approximate
ranges:
 \( f_{\text{BH}}\approx \)   0.1\% to 5\%   \cite{Heger:2002by,Fryer:2011cx}, 
\( \langle m_{\text{prog}} \rangle\approx \)  25 to 40 \( M_\odot \)  
\cite{Woosley:2002zz,Sukhbold:2015wba},  
\( f_{\text{merge}} \approx\)   1\% to 10\%  
\cite{Belczynski:2016obo,Eldridge:2016ymr,Dominik:2012kk}, and  
\( f_{\text{bin}}\approx \)   50\% to 80\%  
\cite{Sana:2012px,Moe:2016tmr,Duchene:2013uma}.
Finally,  introducing  the dimensionless matter density 
parameter at present, namely $\Omega_{m0} = \frac{8 \pi G}{3 H_0^2} \rho_{m0} 
$ (in the following a subscript ``0'' denotes the value of a quantity at 
present),   we can make the substitution $\frac{8 \pi G}{3} \rho_{m} = H_0^2 
\Omega_{m0} (1 + z)^{3}$.

Inserting all the above into \eqref{model 1st Friedmann} 
  we finally 
obtain the modified first Friedmann equation as
 \begin{equation}\label{SFR Hmodel z}
    H^2(z) = H_0^2 \Omega_{m0} (1 + z)^{3} + \frac{\Lambda}{3} - 8 
\tilde{\alpha} C \int_{z_{i}}^z  \frac{ \psi(z)}{ (1 + z)} dz,
\end{equation}
where we have defined for convenience
\begin{equation}
    C \equiv \frac{4 \pi}{3} \frac{ \left( 1 - f_{\text{bin}} 
f_{\text{merge}}\right)  f_{\text{BH}}}{\langle m_{\text{prog}} \rangle}.
\label{Cdef}
\end{equation}

We stress that the astrophysical quantities entering the
effective parameter $C$, namely $f_{BH}$, $f_{merge}$,
$f_{bin}$ and $\langle m_{prog}\rangle$, are currently
subject to significant observational and modeling
uncertainties. Nevertheless, within the present framework
these quantities affect the cosmological evolution only
through the combined parameter combination of relation (\ref{Cdef}), and in 
particular through its multiplication with $\tilde{\alpha}$.
Consequently, the observational analysis effectively
constrains the phenomenological quantity
\begin{equation}
  C_n \equiv \tilde{\alpha}C,
  \label{Cdef2}
\end{equation}
 rather than each
astrophysical parameter individually.

From this perspective, the inferred cosmological
constraints already incorporate the net phenomenological
impact of the underlying astrophysical uncertainties.
However, a more systematic propagation of such
uncertainties, potentially through a hierarchical Bayesian
analysis jointly combining cosmological observations and
black-hole population data, would constitute an important
extension of the present work.

By comparing with the standard form of the Friedmann equation $H^2 = 8 \pi G /3 
\left(\rho_{m} + \rho_{DE}\right)$ we can retrieve the corresponding expression 
for the effective DE density as
\begin{equation}\label{SFR rho z}
     \rho_{DE}(z)  = -\frac{3 \tilde{\alpha} C}{\pi G} \int_{z_{i}}^z  \frac{ 
\psi(z)}{ (1 + z)} dz.
\end{equation}
Fortunately, the integral that appears in the above equations can be evaluated 
analytically with 
the aid of the hypergeometric function ${}_2F_1(a, b; c; z)$ as

\begin{align*}
    \int & \frac{ \psi(z)}{ (1 + z)} dz = 0.005555 \cdot (1 + z)^{2.7} \\ 
\nonumber
    & {}_2F_1\left(0.482143,\, 1.0;\, 1.48214;\, -0.00257378 \cdot (1 + 
z)^{5.6}\right). 
\end{align*}
Moving on, inserting (\ref{SFR rho z}) into the DE equation 
of state parameter $w_{DE} = -1 - \frac{\dot{\rho}_{DE}}{3 H \rho_{DE}}$ we 
obtain
\begin{equation}\label{SFR w z}
    w_{DE}\left(z\right)=-1-\frac{2  \tilde{\alpha} C \psi(z) 
}{\frac{\Lambda}{4} - 6 \tilde{\alpha} C \int_{z_{i}}^z  \frac{ \psi(z)}{ (1 + 
z)} dz}.
\end{equation}

We mention here that  although Einstein-Gauss-Bonnet theory in 4 
dimensions leads to the same field equations with general relativity, 
its incorporation through the gravity-thermodynamics framework yields extra 
terms that arise from the topology and entropy changes on the horizon. Some 
notes are in place here regarding the phenomenology of eq.~\eqref{SFR w z}. 
First of all, if $\tilde{\alpha}$ tends to zero, then there exist an explicit 
$\Lambda CDM$ limit for WGB cosmology. Moreover, as the function $\psi(z)$ goes 
to zero at $z \sim 100$, the integral at eq.~\eqref{SFR Hmodel z} goes to a 
constant value, so  WGB cosmology approaches $\Lambda$CDM with an increased 
(reduced) cosmological constant, given that $\tilde{\alpha}$ has positive 
(negative) sign. From the above, the possibility of affecting the inferred
late-time value of $H_0$ becomes apparent. In particular, negative  
$\tilde{\alpha}$ will reduce the energy density of the Universe today, so 
giving rise to a larger $H_0$ to fit the data. In all cases, the early universe 
behavior will remain identical to $\Lambda$CDM, preserving the thermal history. 
A related approach worth mentioning is the Topological Dark Energy (TDE) model 
\cite{Tsilioukas:2023tdw}, which has recently been shown to provide an 
excellent fit to observations at the background \cite{Anagnostopoulos:2025tax} and the perturbation level \cite{10.1093/ptep/ptag080},\cite{KAVYA2026100621}. Conceptually, 
however, TDE differs from the WGB cosmology in its physical origin: in TDE, dark 
energy arises from spacetime foam, whereas in WGB cosmology it is associated 
with horizon-related effects of black holes. Thus, while both frameworks employ 
similar ingredients—such as the Gauss-Bonnet term and spacetime topology—their 
implementations diverge, leading to distinct cosmological scenarios.
In what follows, we consider two separate model cases, namely $\Lambda = 0$ 
(Model I) and  $\Lambda \neq 0$ (Model II).

\subsubsection{Model I}

One could choose $\Lambda = 0$ and try to assess if the WGB cosmology is able to 
describe fully the DE part of the cosmic budget. This avenue offers the 
intriguing probability of jointly solving the coincidence problem and the 
cosmological constant problem. The coincidence problem is solved as the DE 
energy density is, by construction, proportional to the star formation i.e. 
linked with the matter era. The problem of the cosmological constant 
\cite{Weinberg:1988cp} is solved, as the effective cosmological constant emerges 
from the astrophysical scale, which means that Dark Energy is an emergent 
phenomenon of the late time Universe and goes to zero at large redshifts. A 
related mechanism for simultaneously addressing the coincidence and cosmological 
constant problems has been advanced in \cite{Anagnostopoulos:2018jdq, 
Anagnostopoulos:2022pxa}, grounded in geometrical considerations and 
Asymptotically Safe Quantum Gravity (ASQG) 
\cite{Zarikas:2024chv,Bonanno:2024paf}. The principal distinction between that 
framework and the one under consideration resides in the nature of the operative 
mechanism: the former relies on ASQG in conjunction with a Swiss-cheese 
construction, whereas the latter is founded upon the gravity-thermodynamics 
conjecture and the Wald-Gauss-Bonnet entropy. From an epistemological 
perspective, it lies beyond the present scope to adjudicate between the two 
approaches. From a more phenomenological and practical standpoint, however, an 
appealing aspect of the WGB model is that it involves the same number of free 
parameters as the concordance model, a feature generally regarded as desirable. 
Pursuing this phenomenological line further, in the case where $\Lambda = 0$, 
the equation-of-state parameter becomes strictly phantom - a property that has 
been associated with a possible alleviation of the $H_0$ tension (see, for 
example, \cite{DiValentino:2020naf}). 

Application the normalization condition for the present  $H(z=0) \equiv H_{0}$ 
to \eqref{SFR Hmodel z} results the following expression
\begin{eqnarray}\label{eq:norm}
    \tilde{\alpha} C = -\frac{H_{0}^{2}(1-\Omega_{m0}) }{8 
\int_{z_{i}}^{0}\frac{\psi(z)}{(1+z)}dz }, 
\end{eqnarray}
where it is apparent that the models parameter is a function of the standard 
cosmological parameters $(H_{0}, \Omega_{m0})$.  Thus, the Model I has the same 
number of parameters as the concordance model. Substituting \eqref{eq:norm} to 
\eqref{SFR Hmodel z} we obtain the dimensionless Hubble rate
\begin{equation}\label{eq:dimnlss Friedmann}
    E(z)=\left(\Omega_{m0}(1+z)^{3} + (1-\Omega_{m0}) \frac{ \int_{z_{i}}^{z} 
\psi(z)/(1+z)dz}{\int_{z_{i}}^{0} \psi(z)/(1+z)dz}\right)^{1/2}
\end{equation}
At $z=0$, as the ratio with the integrals becomes unity, Model I coincides with 
$\Lambda$CDM. However, in the limit of $z \rightarrow z_i$, Model I goes to bare 
GR, without cosmological constant. For small redshifts ($z << z_i$), the ratio  
with the integrals takes values in the range $(0,1]$, thus exhibits a 
diminishing value of $\Lambda$. Note that Model I does not posses explicit 
$\Lambda$CDM limit.

\subsubsection{Model II}
Requiring an explicit $\Lambda$CDM limit for the WGB scenario, we consider the 
case where $\Lambda \neq 0$. Here the normalization condition for the present  
$H(z=0) \equiv H_{0}$ to \eqref{SFR Hmodel z} leads to an expression for the 
cosmological constant as a function of the $\tilde{a}C$ parameter
\begin{eqnarray}\label{eq:norm}
    \Lambda = 3(1-\Omega_{m0})H_0^2 
-24\tilde{a}C\int_{0}^{z_{i}}\frac{\psi(z)}{(1+z)}dz , 
\end{eqnarray}
A subsequent question has to  do with the whether it is possible to obtain a 
negative effective cosmological constant, a situation of particular 
observational interest, i.e. \cite{Sen:2021wld,Luu:2025dax}. 
The latter corresponds to
\begin{equation}
\tilde{a}C \geq \frac{(1-\Omega_{m0})H_0^2}{8I(z)}
\end{equation}
where $I(z)$ denotes the integral appearing on \eqref{eq:modelII-Erate}. As the 
right side of the inequality is always positive, negative $\tilde{a}$ 
corresponds to positive cosmological constant, while positive $\tilde{a}$ could 
in principle allow for negative effective cosmological constant.
The dimensionless Hubble rate reads as follows
\begin{equation}
\label{eq:modelII-Erate}
    E^2(z) = \Omega_{m0}(1+z)^3  + 1 - \Omega_{m0} - \frac{8 
\tilde{a}C}{H_0^2}\int_{0}^{z}\frac{ \psi(z)}{ (1 + z)}
\end{equation}
As expected, there exists an explicit $\Lambda$CDM limit for $\tilde{a} =0$. 
Moreover, for positive $\tilde{a}$ the WGB scenario provides a reduced 
cosmological constant. 
The equation of state reads as
\begin{equation}
       w_{DE}\left(z\right)=-1-\frac{8  \tilde{\alpha} C \psi(z) 
}{3(1-\Omega_{m0})H_0^2 + 6 \tilde{\alpha} C \int_{0}^z  \frac{ \psi(z)}{ (1 + 
z)} dz}.
\end{equation}
It is apparent that for positive (negative) $\tilde{a}C$ we have phantom 
(quintessence) behavior.

\subsection{Perturbation analysis}

At this point, we extend the previous work of \cite{Tsilioukas:2024seh}, 
developing the perturbation analysis for the WGB cosmology in the context of the 
effective fluid approach. There, the perturbations of matter, Dark Energy and of 
the gravitational scalar potential form a system of coupled odes, where the DE 
is modeled by a fluid with equation of state parameter $w_{d}$ and effective 
sound speed $c_e$ \cite{Hu:1998kj,Nesseris:2022hhc}. In particular, one can 
effectively describe all additional terms in the modified Einstein equations as 
to be produced by an extra component of $T_{\mu \nu}$.  We use the formalism of  
\cite{Mehrabi:2015hva}, where a transformation from the conformal time to 
redshift results the following equations:
\begin{align}\label{eq:pert_phi}
    \Phi''(z)&=\frac{1}{E^{2}(z)}\Bigg[\frac{3c_e 
\delta_{d}(z)E^2(z)\Omega_{d}(z)}{2(z+1)^2}-\bigg(E(z)E'(z)\nonumber\\
    &-\frac{3E(z)^2}{z+1}\bigg)\Phi'(z)
    -\Phi(z)\left(\frac{3E^2(z)}{(z+1)^2}-\frac{2E(z)E'(z)}{z+1}\right)\Bigg]
\end{align}
The function $\Omega_{d}(z)$ can be written as follows:
\begin{equation}
    \Omega_{d}(z) = E^2(z) - \Omega_{m0}(1+z)^3.
\end{equation}
The equation for the matter over-density evolution:
\begin{eqnarray}\label{eq:overd_eq_m_z}
    \frac{d^2\delta_{m}(z)}{dz^2} &=&-\frac{\left[2(z+1)^3-(z+1)^2 
A_{m}(z)\right]}{(z+1)^4}\delta_{m}'(z) \nonumber \\ &&- 
\frac{B_{m}(z)\delta_{m}(z)+S_{m}(z)}{(z+1)^4},
\end{eqnarray}
where the coefficients are:
\begin{eqnarray}\label{eq:coeff_m_z}
A_{\rm m}(z) & = & 
\frac{3}{2}\,(1+z)\,\bigl[1-\Omega_{d}(z)\,w_{d}(z)\bigr],\nonumber\\[4pt]
B_{\rm m}(z) & = & 0,\nonumber\\[4pt]
S_{\rm m}(z) & = & 3\,(1+z)^{4}\,\Phi''(z)\nonumber \\
                   + &&\frac{3}{2}\,(1+z)^{3}\,\bigl[1+3\Omega_{d} 
(z)\,w_{d}(z)\bigr]\,\Phi'(z)\nonumber\\[4pt]                  
-&&\left(\frac{k}{H_{0}}\right)^{2}\frac{(1+z)^{4}}{E(z)^{2}}\,\Phi(z).\nonumber
\end{eqnarray}
The equation for the DE over-density evolution: 
\begin{eqnarray}\label{eq:delta_ddelta_DE}
    \frac{d^2\delta_{d}(z)}{dz^2} &=&\frac{-\left[2(z+1)^3- (z+1)^2 
A_{d}(z)\right]}{(z+1)^4}\delta_{d}'(z) \nonumber \\ &&  - 
\frac{B_{d}(z)\delta_{d}(z)+S_{d}(z)}{(z+1)^4},
\end{eqnarray}
and the coefficients are:
\begin{eqnarray}\label{eq:coeff_d_z}
A_{\rm d}(z) & = &
(1+z)\Bigl[\tfrac32\bigl(1-\Omega_{d}(z)w_{d}(z)\bigr)
            +3c_{a}(z)-6w_{d}(z)\Bigr],\nonumber\\[6pt]
B_{\rm d}(z) & = &
\Biggl\{
3\bigl(c_{e}-w_{d}(z)\bigr)
  \Bigl[\tfrac12 - w_{d}(z)\left[\tfrac32\Omega_{d}(z)
        +3\right]\nonumber\\
&&
+3c_{a}(z)-3c_{e}\Bigr]+\left(\frac{k}{H_{0}}\right)^{2}\frac{(1+z)^{2}}{E(z)^{2
}}\,c_{e}\nonumber\\
&&
+3(1+z)\,\frac{dw_{d}}{dz}\Biggr\}(1+z)^{2}\,\nonumber\\
[6pt]
S_{\rm d}(z) & = &
(1+w_{d}(z))
\Bigl[\,3(1+z)^{4}\,\Phi''(z)\nonumber\\
&& 
+\,(1+z)^{3}\!\Bigl(\tfrac32
       +\tfrac92\,\Omega_{d}(z)w_{d}(z)+9c_{a}(z)\Bigr)\,\Phi'(z)\nonumber\\
&& 
-\left(\frac{k}{H_{0}}\right)^{2}\frac{(1+z)^{4}}{E(z)^{2}}\,\Phi(z)\Bigr]
\nonumber\\
&&\;+\;\frac{3(1+z)^{4}}{1+w_{d}}\,\Phi'(z)\,\frac{dw_{d}}{dz}.\nonumber
\end{eqnarray}

The initial conditions are the following \cite{Mehrabi:2015hva}:

\begin{equation}
\delta_{\rm m,i}
\;=\;
-\,2\,\phi_{i}\left[
1+\left(\frac{k}{H_{0}}\right)^{2}\frac{(1+z_{i})^{2}}{3\,E(z_{i})^{2}}
\right]\;,
\end{equation}

\begin{equation}
\left.\frac{d\delta_{\rm m}}{dz}\right|_{z=z_i}
= \frac{2}{3}\,
  \left(\frac{k}{H_{0}}\right)^{2}\frac{1}{E(z_i)^{2}}\,
  \phi_{i}\;,
\end{equation}

\begin{equation}
\delta_{\rm d,i}
\;=\;
\bigl[\,1+w_{d}(z_{i})\bigr]\,
\delta_{\rm m,i}\;,
\end{equation}

\begin{align}
\left.\frac{d\delta_{\rm d}}{dz}\right|_{z=z_i}
=&
\bigl[1+w_{d}(z_i)\bigr]\,
\left(\frac{k}{H_{0}}\right)^{2}\frac{2}{3\,E(z_i)^{2}\,}\,
\phi_{i}\nonumber\\
\;+\;
&\left.\frac{dw_{d}}{dz}\right|_{z=z_i}\,
\delta_{\rm m,i}\;.
\end{align}
Up to now, we have only used the approximation $\frac{\delta 
p_d}{\delta\rho_d}=c_e^{2}=c_{e}$. The above system of equations, along with the 
corresponding initial conditions can be solved numerically. Furthermore, after 
extracting the solution for $\delta_m (z)$ one can  
calculate the important physical observable 
\begin{equation}
    f\sigma 8 \equiv f(z)\sigma(z),
\end{equation}
where $f(z) := -\frac{d ln \delta_m (z)}{d ln z}$ and  $\sigma (z) := \sigma_8 
\frac{\delta_m (z)}{\delta_m (0)}$ \cite{Abdalla:2022yfr}.

We apply the quasi-static approximation, in the sense that the time derivatives 
of the fields are considered negligible with regard to the spatial derivatives 
(i.e. terms where $k^2$ appears), then \eqref{eq:overd_eq_m_z} with 
\eqref{eq:coeff_m_z} becomes

\begin{eqnarray}
    \delta_{m}''(z) & + & \left(\frac{2}{1+z}-\frac{3}{2(1+z)}[1-\Omega_{d}(z) 
w_d(z)] \right)\delta_{m}'(z)  \nonumber \\  &\simeq &\frac{k^2}{H^2}\Phi(z)
\end{eqnarray}
and further, from the Poisson equation in the sub-horizon approximation, 
(Appendix B in \cite{Mehrabi:2015hva} )
\begin{equation}
    k^2\Phi(z) \simeq 4\pi Ga^2\left[\rho_{m}\delta_m(z) + (1 + 3 c_{e}) 
\rho_{d}\delta_d(z)\right]
\end{equation}
which allows us to write

\begin{eqnarray}
\label{eq:delta-approx-fin}
     \delta_{m}''(z) + \frac{1+3\Omega_{d}(z) w_d(z)}{2 (z+1)}\delta_{m}'(z) 
&\simeq & \nonumber \\   \frac{3}{2}\frac{\Omega_{m0} (1+z)}{E^2(z)} 
\frac{G_{eff}(z)}{G_N}\delta_m(z)
\end{eqnarray}

where
\begin{equation}\label{eq:G_eff}
    \frac{G_{eff}(z)}{G_N} = \left(1 + (1 + 3 c_{e}) 
\frac{\Omega_d(z)\delta_d(z)}{\Omega_{m0}(1+z)^3\delta_m(z)}\right)
\end{equation}

In the limit of $w_d \rightarrow-1$, $\delta_{d} \rightarrow 0$ and $c_{e} 
\rightarrow 0$, it is easy to show that equation \eqref{eq:delta-approx-fin}, 
reduces to the standard form (i.e eq. 2.2 of \cite{Kazantzidis:2019nuh}).

\section{Observational constraints}
\label{obs-constr}
\subsection{Data and Methodology}
In order to assess the observational effectiveness of the WGB cosmology, we 
confront both the above models with observational data from Supernovae Type Ia 
(SNIa), Cosmic Chronometers (CC) measurements and Baryonic Acoustic Oscillations 
(BAO). 

Concerning the SNIa data, we utilize the full Pantheon+/SH0ES sample 
\cite{Scolnic:2021amr,Pan-STARRS1:2017jku}, with data points within the redshift 
range $0.001 \lesssim z 
\lesssim 2.27$. The   chi-square function is given by 
$
          \chi^{2}_{SN Ia}\left(\phi^{\nu}\right)={\bf \mu}_{\text{SNIa}}\,
          {\bf C}_{\text{SNIa},\text{cov}}^{-1}\,{\bf \mu}_{\text{SNIa}}^{T}\,,
$
where    ${\bf 
\mu}_{\text{\text{SNIa}}}=\{\mu_{1}-\mu_{\text{th}}(z_{1},\phi^{\nu})\,,\, 
...\,, \,   \mu_{N}-\mu_{\text{th}}(z_{N},\phi^{\nu})\}$, where 
$\phi^{\lambda}$ is the statistical vector  with the free parameters. Moreover, 
 the distance modulus is   $\mu_{i} = \mu_{B,i}-\mathcal{M}$, with 
$\mu_{B,i}$ the apparent magnitude at maximum brightness in the rest frame of 
  $z_{i}$. The  parameter $\mathcal{M}$
   accounts for the dependence of the observed 
distance modulus, $\mu_{obs}$, on $H_0$, and on the fiducial 
cosmological model employed . 
Lastly, the theoretical distance modulus is 
\begin{equation}
    \mu_{\text{th}} = 5\log\left[\frac{d_{L}(z)}{\text{Mpc}}\right] + 25,
\end{equation}
where \begin{equation}
d_L(z) = c(1+z)\int_{0}^{z}\frac{dx}{H(x,\phi^{\nu})}
\end{equation} is
 the luminosity distance,  assuming  spatially flat Universe. 

 Concerning the Cosmic Chronometers   we employ the latest compilation of the 
$H(z)$ dataset, as 
presented in \cite{Magana:2017nfs}. Our analysis
incorporates a total of $N=22$ measurements   data 
within 
$0.07 \lesssim z \lesssim 2.0$. In this case the corresponding $\chi^2_{H}$ 
function is expressed as 
$
          \chi^{2}_{H}\left(\phi^{\nu}\right)={\bf \cal H}\,
          {\bf C}_{H,\text{cov}}^{-1}\,{\bf \cal H}^{T}\,,
  $
with ${\bf \cal H 
}=\{H_{1}-H_{0}E(z_{1},\phi^{\nu})\,,\,...\,,\,
          H_{N}-H_{0}E(z_{N},\phi^{\nu})\}$ and where $H_{i}$ are the observed 
Hubble   values at    $z_{i}$ ($i=1,...,N$). 

Finally, we utilize the DESI BAO measurements \cite{DESI:2024mwx}. BAO are 
observed as periodic variations in the density of visible baryonic matter and 
function as a standard cosmological ruler, set by the sound horizon at the drag 
epoch. The sound horizon, represents the maximum distance that sound waves could 
travel before baryons decoupled in the early universe, leaving a characteristic 
scale in the matter distribution. For the case of the concordance model, the 
latter is given by $r_d  = \int^{\infty}_{z_d}c_s(z)/H(z)dz $, where $z_d$ is 
the redshift of the drag epoch and $c_s$ the sound speed. As customary (see 
\cite{Anagnostopoulos:2018jdq} and references therein), we
leave $r_d$ as a free parameter.

The BAO measurements used in this study are obtained from various samples: 
The Bright Galaxy Sample (BGS, 
$0.1 < z < 0.4$), the Luminous Red Galaxy Sample (LRG, $0.4 < z < 0.6$ and $0.6 
< z < 0.8$), the Emission Line Galaxy Sample (ELG, $1.1 < z < 1.6$), the 
combined LRG and ELG Sample (LRG+ELG, $0.8 < z < 1.1$), the Quasar Sample (QSO, 
$0.8 < z < 2.1$) and the Lyman-$\alpha$ Forest Sample (Ly$\alpha$, $1.77 < z < 
4.16$). The chi-squared statistic used to fit the BAO data is
$\chi^2_{\text{BAO}}= (\Delta {X})^T 
\mathbf{C_{BAO}}^{-1} \Delta {X}$
where $\Delta {X} =  {x^{obs}} -  {x^{th}}$ and 
$\mathbf{C_{BAO}}^{-1}$ the inverse covariance matrix.

Our analysis is fully Bayesian, based on the likelihood function
$\mathcal{L}_\mathrm{tot} \sim \exp\left(-\chi^2_\mathrm{tot}/2\right),$
where the total chi-squared, $
\chi^2_\mathrm{tot} = \chi^2_\mathrm{H} + \chi^2_\mathrm{SNIa} + 
\chi^2_\mathrm{BAO}
$, each of which is presented in the previous paragraphs. According to 
\cite{Leizerovich:2023qqt}, the considered datasets are compatible, however in 
order to assess a possible tension between the datasets, we consider all 
possible combinations, namely CC/Pantheon+/SH0ES/BAOs, CC/Pantheon+/SH0ES, 
CC/BAOs. 

The parameter space explored in this work is described by the vector $
{\phi}^\nu = \{ H_0, \Omega_{m0}, r_d, C_n \}$,
where $H_0$ is the Hubble constant, $\Omega_{m0}$ is the present value of 
matter density parameter, $r_d$ is the sound horizon at the drag epoch, and $C_n 
\equiv \tilde{a} \, C$, defined in  (\ref{Cdef2}), is the parameter of the 
model. In Tab. \ref{tab:priors} we show the flat priors adopted for the 
aforementioned parameters.
\begin{table}[h!]
\centering
\label{tab:priors}
\begin{tabular}{lcc}
\hline
\hline
\textbf{Parameters} & \textbf{Min} & \textbf{Max} \\
\hline
\hline
$H_0$ & 40 & 120 \\
$\Omega_{m0}$ & 0.1 & 0.9 \\
$r_d$ & 110 & 200 \\
$C_n$ & $-1 \cdot 10^5$ & $1 \cdot 10^5$ \\
\hline
\hline
\end{tabular}
\caption{Priors for the cosmological parameters used in the analysis.}
\end{table}

We use the Cobaya’s Monte Carlo Markov Chain
(MCMC) sampler \cite{Torrado:2020dgo} to obtain the posterior distributions of 
the cosmological parameters of our model. The sampling is performed with 4 
parallel chains, each evolving for $\approx 4080$ accepted steps. Convergence is 
verified using the Gelman-Rubin criterion, demanding $R-1 < 0.01$.

\subsection{Information Criteria}

To evaluate the statistical performance of our model compared to $\Lambda$CDM, 
we employ three widely used information criteria: the Akaike Information 
Criterion (AIC), 
the Bayesian Information Criterion (BIC) and the Deviance Information Criterion 
(DIC) \cite{Liddle:2007fy}, following the standard Jeffreys scale 
\cite{Kass:1995loi}.
These criteria serve as important tools for model selection.

The AIC is defined as
\begin{equation}
    \mathrm{AIC} = -2\ln L_{\max} + 2k,
\end{equation}
where $L_{\max}$ is the maximum likelihood of the model and $k$ is the number of 
free parameters.  

The BIC is given by
\begin{equation}
    \mathrm{BIC} = -2\ln L_{\max} + k \ln N,
\end{equation}
where $N$ denotes the number of data points. 
While AIC tends to favor models with better fits, BIC introduces a stronger 
penalty for additional parameters, 
especially for large datasets, making it a more conservative criterion.

Finally, the DIC combines features of both AIC and BIC, and is defined as
\begin{equation}
    \mathrm{DIC} = D(\hat{\phi}) + 2C_B,
\end{equation}
where $D(\phi) = -2\ln L(\phi)$ is the Bayesian deviance, $\hat{\phi}$ denotes 
the posterior mean of the parameters, 
and $C_B$ represents the Bayesian complexity, which quantifies the effective 
number of parameters constrained by the data.

For model comparison, one typically evaluates the relative differences 
\begin{equation}
    \Delta \mathrm{DIC} = \mathrm{DIC}_{\text{model}} - \mathrm{DIC}_{\min},
\end{equation}
where $\mathrm{DIC}_{\min}$ is the lowest value among the models considered. 
According to the Jeffreys’ scale, a difference $\Delta \mathrm{DIC} \leq 2$ 
indicates statistical equivalence with the best model, 
$2 < \Delta \mathrm{DIC} < 6$ suggests moderate tension between the model at 
hand and the best model, while larger values point to strong evidence against 
the model.

\subsection{Results \& Discussion}
\begin{table*}[htb]
\centering
\renewcommand{\arraystretch}{1.2}
\setlength{\tabcolsep}{6pt}
\begin{tabular}{llcccccc}
\hline\hline
Model & $\Omega_{m0}$ & $C_n$ & $h$ & $r_d$ & $\chi^2_{\rm min}$ & $\chi^2_{\rm 
min}/\mathrm{dof}$ \\
\hline
\multicolumn{7}{c}{\textbf{CC/Pantheon+/SH0ES/BAOs}} \\
\hline
Model I           & $0.320\pm0.011$ & $-$                     & $0.723\pm0.009$  
       & $140.0\pm2.1$      & $1494.11$      & $0.85$ \\
Model II           & $0.287\pm0.026$ & $-3298.4\pm4636.9$                    & 
$0.723\pm0.008$         & $139.7\pm1.9$      & $1492.69$      & $0.85$ \\
$\Lambda$CDM  & $0.303\pm0.011$ & $-$      
& $0.724\pm 0.008$      & $139.9\pm 1.9$     & $1492.05$   &$0.85$ \\
\hline
\multicolumn{7}{c}{\textbf{Pantheon+/SH0ES/BAOs}} \\
\hline
Model I           & $0.327\pm0.012$ & $-$                     & $0.738\pm 0.011$ 
        & $136.7\pm 2.3$      & $1472.27$      & $0.85$ \\
Model II           & $0.288\pm0.028$   & $-4639.1\pm5314.6$                   & 
$0.736\pm 0.011$         & $136.2\pm 2.3$      & $1470.52$      & $0.85$ \\
$\Lambda$CDM  & $0.311\pm 0.012$ & $-$                        & $0.737\pm 0.010$ 
        & $136.6\pm2.3 $      & $1470.56$      & $0.85$ \\
\hline
\multicolumn{7}{c}{\textbf{CC/BAOs}} \\
\hline
Model I           & $0.309\pm0.014$ &$-$                     & $0.695\pm0.018$   
      & $147.3\pm3.6$      & $30.74$      & $0.57$ \\
Model II           & $0.301\pm0.028$   & $1497.2\pm6415.1$                  & 
$0.695\pm0.019$         & $147.0\pm3.4$      & $31.35$      & $0.59$ \\
$\Lambda$CDM  & $0.296\pm0.014$  & $-$       &           $0.691\pm0.017$         
& $147.4\pm 3.5$      & $30.50$      & $0.56$ \\
\hline\hline
\end{tabular}
\caption{Parameter estimation results for the WGB and $\Lambda$CDM models along 
with the $\chi^2_{min}$ values and the 1-$\sigma$ interval.} 
\label{params_wgb_lcdm}
\end{table*}

\begin{table}[h]
\centering
\setlength{\tabcolsep}{2pt}
\label{tab:info_criteria_wgb_lcdm}
\begin{tabular}{llcccccc}
\hline\hline
Model & AIC & $\Delta$AIC & BIC & $\Delta$BIC & DIC & $\Delta$DIC \\
\hline
\multicolumn{7}{c}{\textbf{CC/Pantheon+/SH0ES/BAOs}} \\
\hline
Model I          & $1500.11$ & $2.06$ & $1516.53$ & $2.06$ & $1500.10$ & $2.11$ 
\\
Model II          & $1500.69$ & $2.64$ & $1522.58$ & $8.11$ & $1500.55$ & $2.56$ 
\\
$\Lambda$CDM & $1498.05$ & $0.0$ & $1514.47$ & $0.0$ & $1497.99$ & $0.0$ \\
\hline
\multicolumn{7}{c}{\textbf{Pantheon+/SH0ES/BAOs}} \\
\hline
Model I         & $1478.27$ & $1.71$ & $1494.65$ & $1.71$ & $1478.16$ & $1.63$ 
\\
Model II          & $1478.52$ & $1.96$ & $1500.35$ & $7.41$ & $1478.51$ & $1.98$ 
\\
$\Lambda$CDM & $1476.56$ & $0.0$ & $1492.94$ & $0.0$ & $ 1476.53$ & $0.0$ \\
\hline
\multicolumn{7}{c}{\textbf{CC/BAOs}} \\
\hline
Model I          & $36.74$ & $0.24$ & $42.87$ & $0.24$ & $36.71$ & $0.32$ \\
Model II          & $39.35$ & $2.85$ & $47.52$ & $4.89$ & $39.31$ & $2.92$ \\
$\Lambda$CDM & $ 36.50$ & $0.0$ & $42.63$ & $0.0$ & $36.39$ & $0.0$ \\
\hline\hline
\end{tabular}
\caption{The information criteria AIC, BIC, and DIC for both WGB models and 
$\Lambda$CDM models, 
alongside the corresponding differences.}
\end{table}

\begin{figure*}[h!]
    \centering
\includegraphics[scale=0.55]{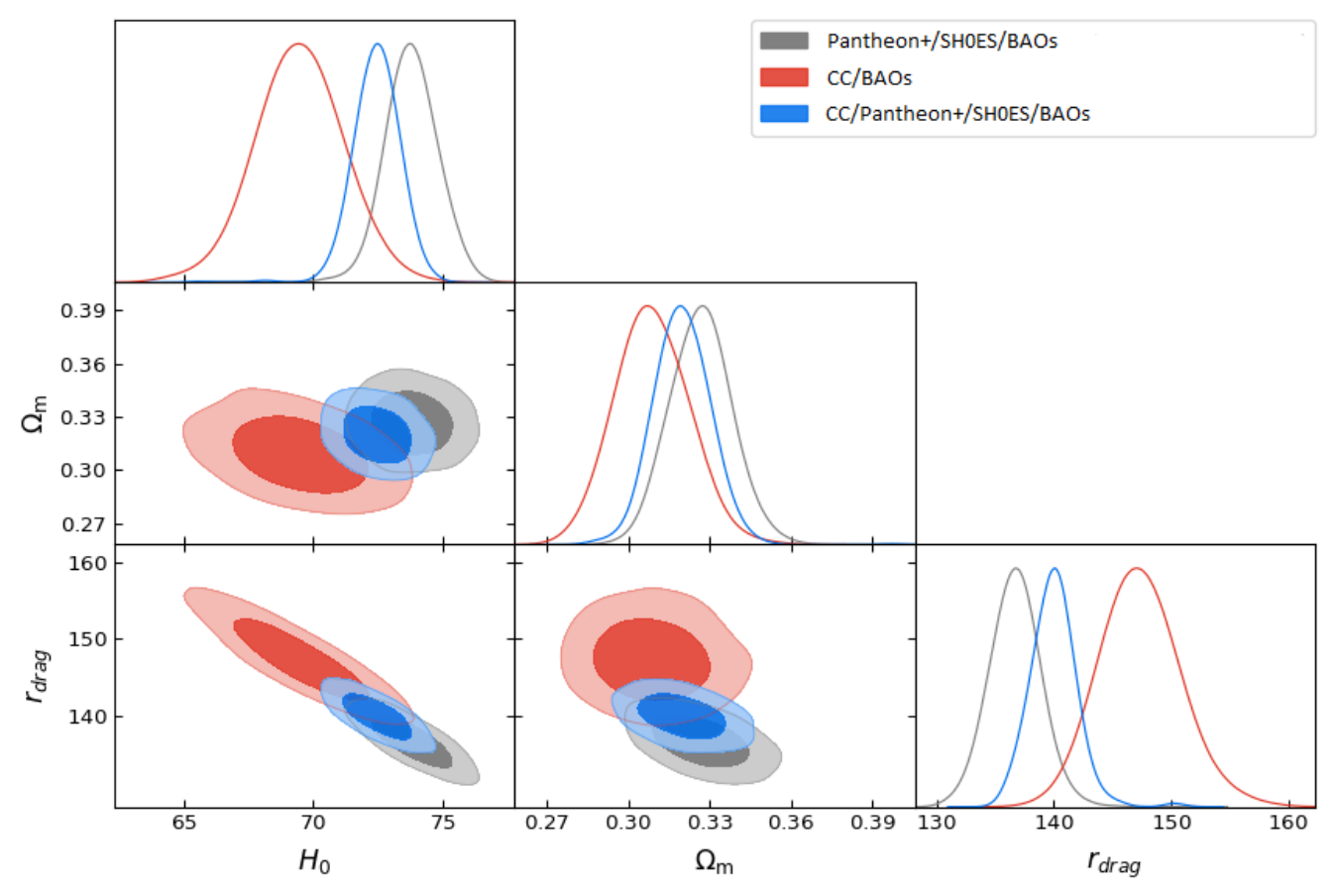}
\includegraphics[scale=0.75]{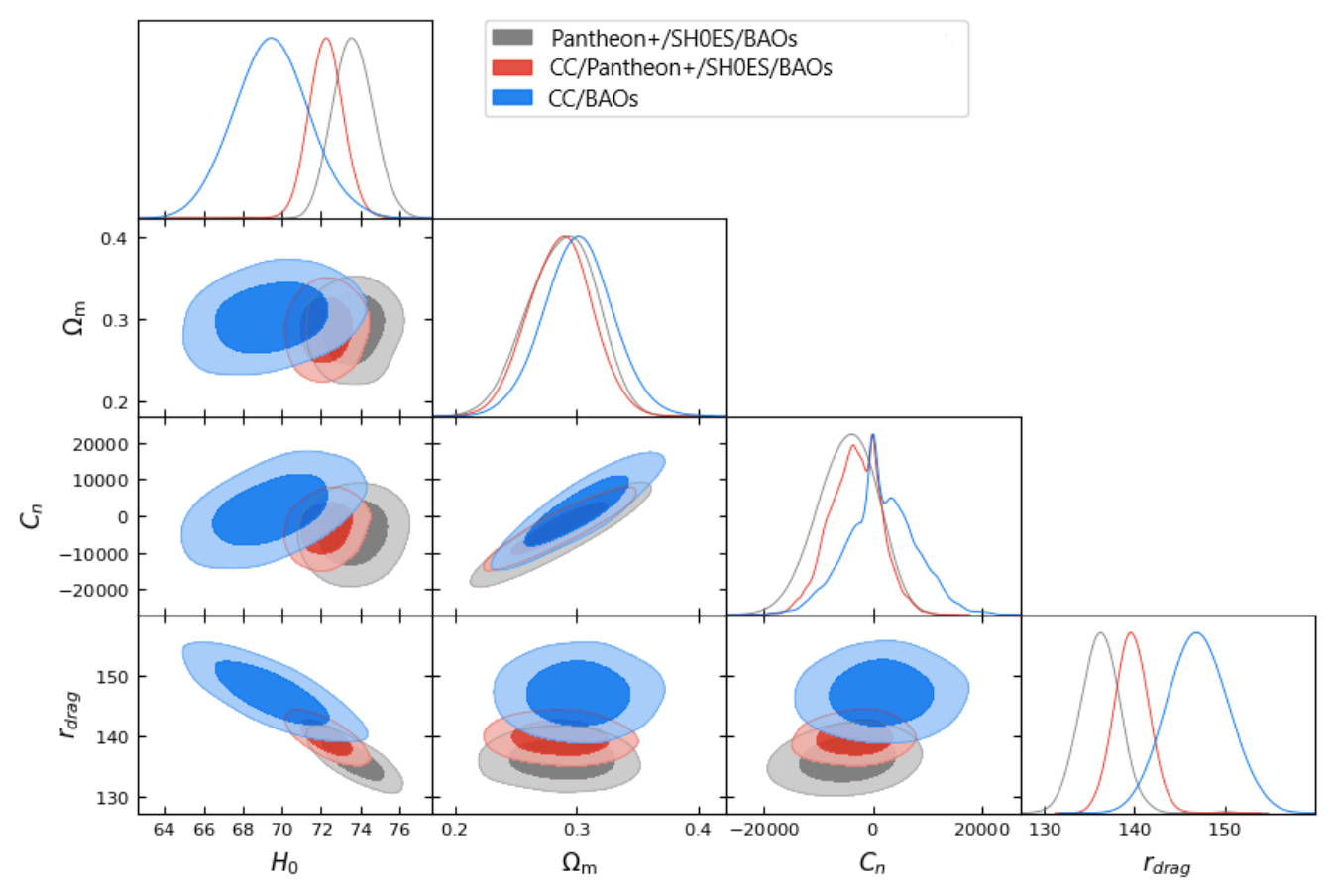}
    \caption{Two-dimensional posterior distributions for the free parameters of 
both WGB models, for all dataset combinations. The contours correspond to the 
$68\%$ and $95\%$ confidence levels, defined in a mode-independent way via 
quantiles.}
    \label{contplot}
\end{figure*}

of 
\begin{figure*}[!htbp]
    \centering
\includegraphics[scale=0.5]{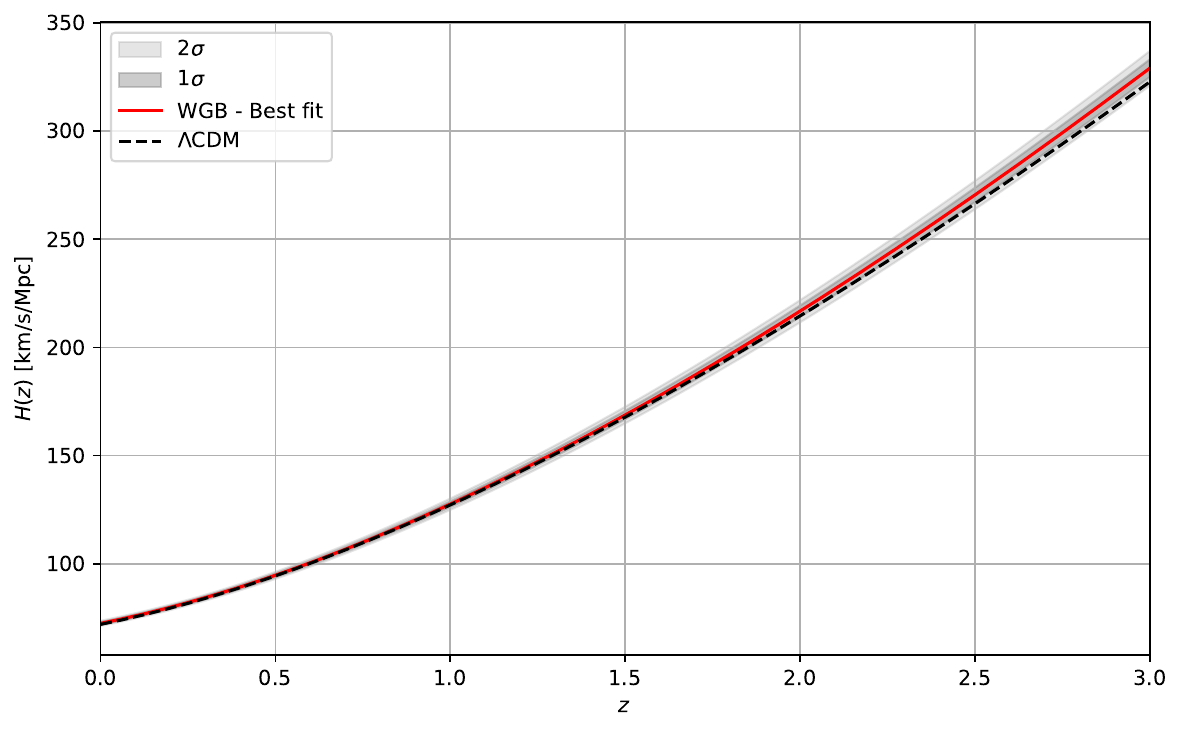}
\includegraphics[scale=0.5]{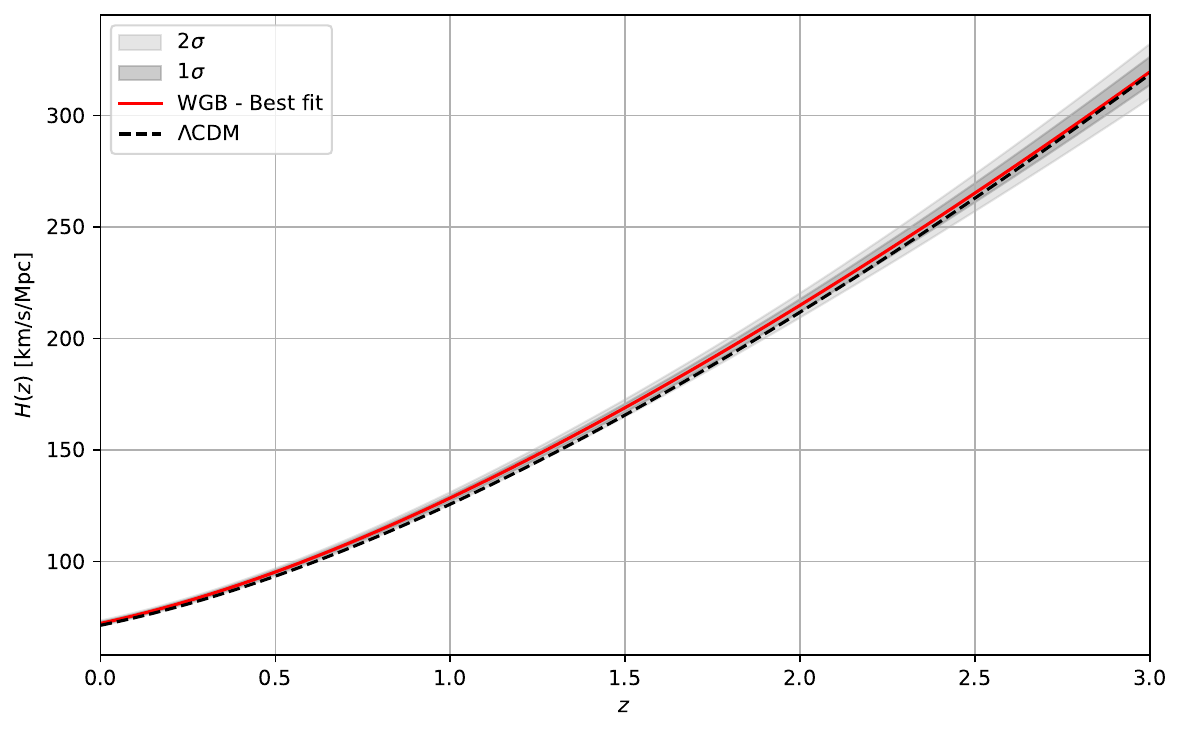}
    \caption{Reconstruction of the Hubble parameter  for WGB cosmology, using 
the best fit values from CC/Pantheon+/SH0ES/BAOs dataset. The grey shaded areas, 
correspond to $1\sigma$ (deep
grey) and $2\sigma$ (light grey) regions. The red line corresponds to the best 
fit parameter
values for the WGB scenario and the black dashed line to the Hubble rate for the 
$\Lambda$CDM scenario. \textit{Upper panel:} Model I \textit{Lower panel:} Model 
II.}
    \label{fig:hubblerec}
\end{figure*}

\begin{figure}
    \centering
\includegraphics[scale=0.6]{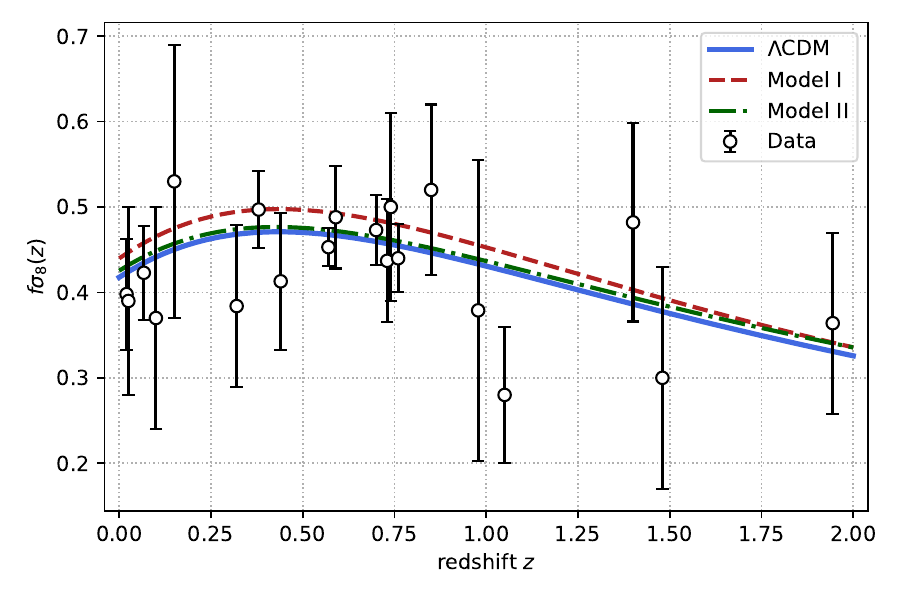}
    \caption{\textit{Evolution of $f\sigma_8$ in  Model I is represented  (red 
dashed line) and Model II (green dash-dot-line) using the best fit parameters 
from the complete dataset CC/Pantheon+/SH0ES/BAOs, considering for various 
values of the effective speed sound of DE $c_e$ with each value noted with a 
distinct color. The $f\sigma_8$ for $\Lambda$CDM appears with blue line. In the 
above calculations we have used $\sigma_8=0.81$ \cite{Stolzner:2025htz}. The 
data points with their error-bars have been obtained from 
\cite{Avila:2022xad}.}}
    \label{fig:fs8_c_e}
\end{figure}

\begin{figure}
    \centering
\includegraphics[scale=0.45]{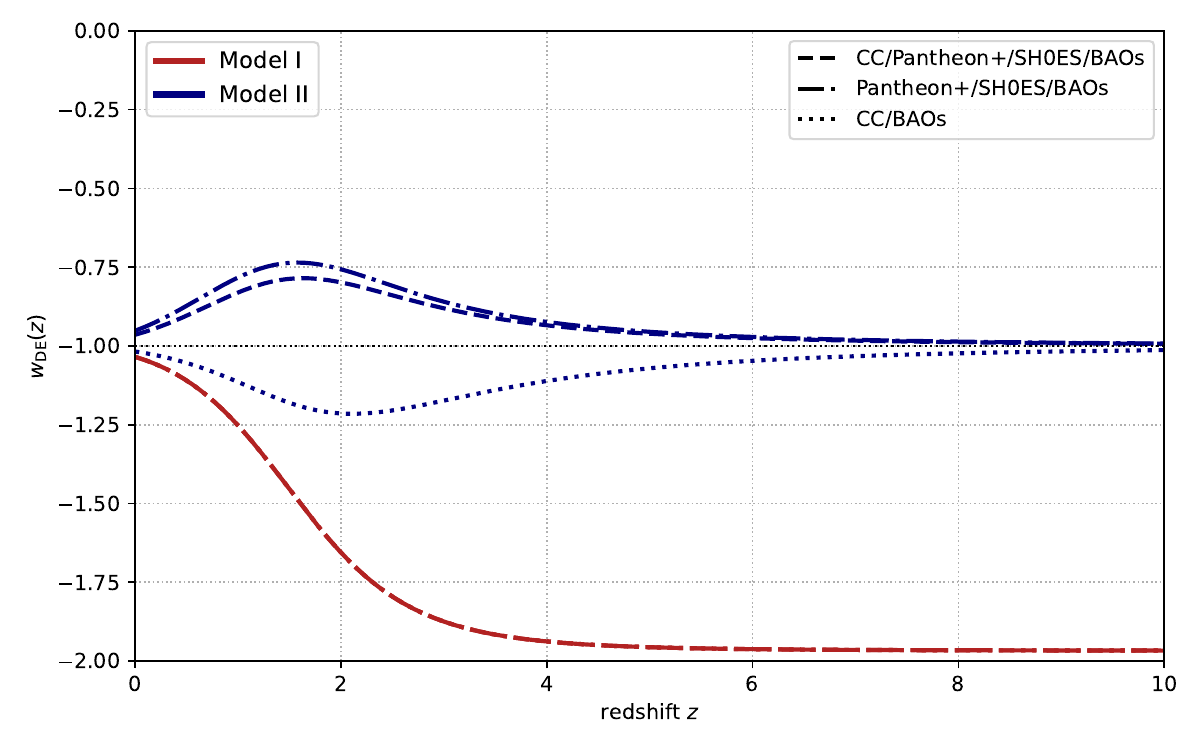}
    \caption{\textit{The evolution of the DE equation of state parameter in  
Model I is represented  (red line) and Model II (blue line), using the best fit 
parameters from the datasets CC/Pantheon+/SH0ES/BAOs (dashed), 
Pantheon+/SH0ES/BAOs (dashed - dotted) and CC/BAOs (dotted).}}
    \label{fig:w_DE}
\end{figure}

\begin{figure}
    \centering
 \includegraphics[scale=0.5]{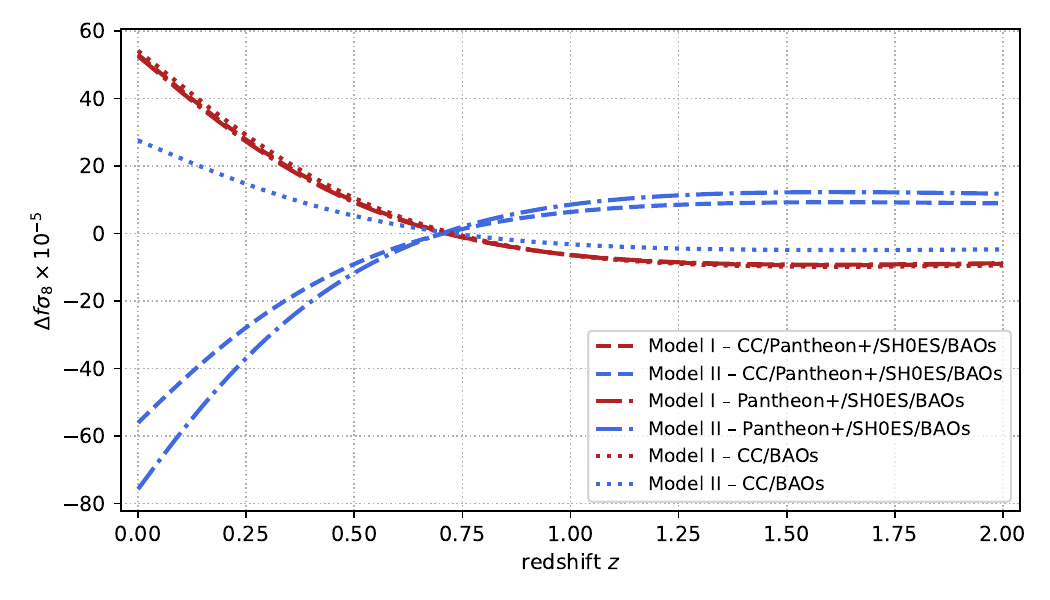}
     \caption{\textit{ The difference $\Delta f\sigma_{8} = f\sigma_{8}(c_e=1) - 
f\sigma_{8}(c_e=0)$, with respect to the acoustic speed of DE $c_e$, in  
Wald-Gauss-Bonnet cosmology, Model I (red) and Model II (blue), for the best fit 
parameters in the following datasets: CC/Pantheon+/SH0ES/BAOs  (dashed), 
Pantheon+/SH0ES/BAOs (dashed - dotted), CC/BAOs  (dotted). In the above 
calculations we have used $\sigma_8=0.81$ \cite{Stolzner:2025htz}.}}
    \label{fig:Delta_fs8}
\end{figure}

\begin{figure}
    \centering
\includegraphics[scale=0.4]{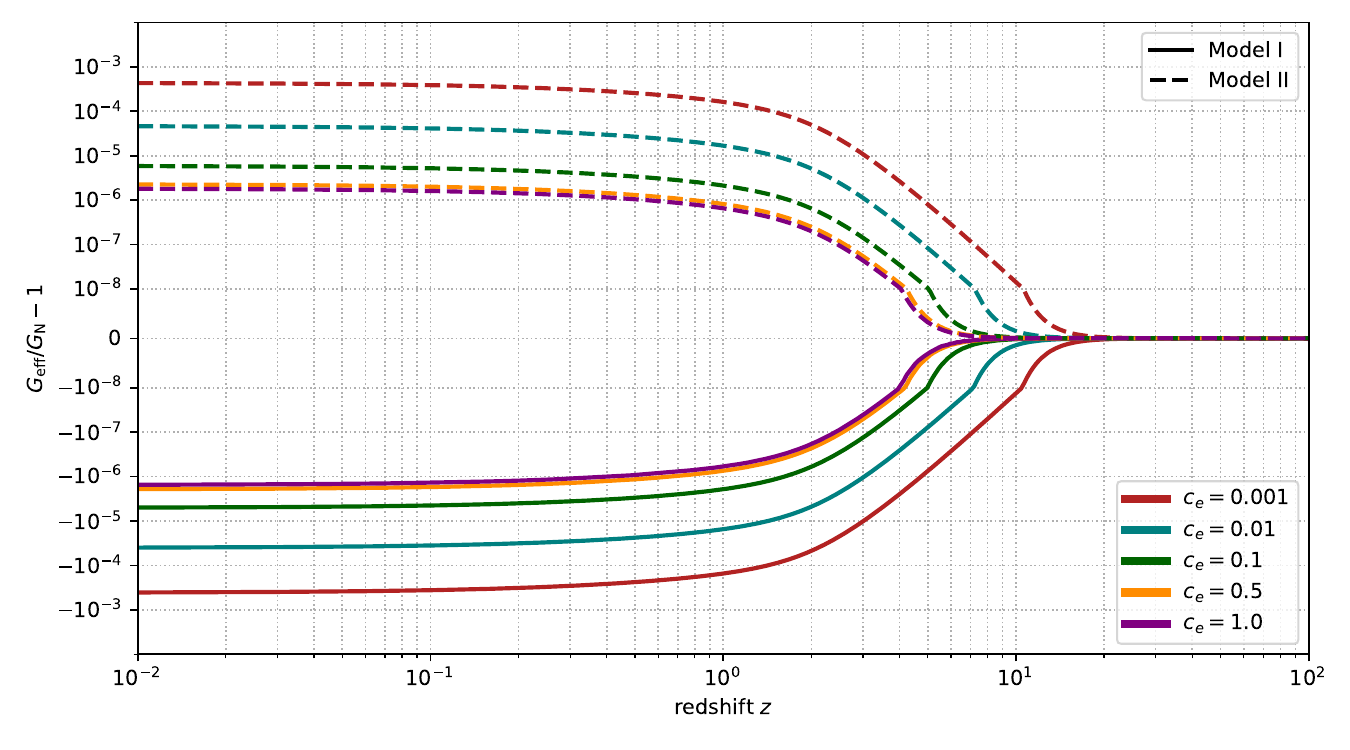}
    \caption{\textit{Evolution of the quantity $G_{eff}/G_{N}-1$ in  
Wald-Gauss-Bonnet cosmology for the best fit parameters of the 
CC/Pantheon+/SH0ES/BAOs and for various values of the effective sound speed of 
DE $c_e$, each value noted with a distinct color. Dashed lines correspond to 
Model II , while continuous lines to Model I.}}
    \label{fig:Geff_c_e}
\end{figure}

In this section, we summarize the results for the posterior parameter values in 
Tab.  \ref{params_wgb_lcdm}, while in Tab. \ref{tab:info_criteria_wgb_lcdm} we 
present the corresponding values for the model selection criteria. The posterior 
distributions of both Model I and Model II parameters against 
CC/Pantheon+/SH0ES/BAOs, Pantheon+/SH0ES/BAOs and  CC/BAOs datasets are 
illustrated in Fig. \ref{contplot} as iso-likelihood contours on two-dimensional 
subspaces of the parameter space (triangle plot). 
As we observe in Table \ref{params_wgb_lcdm},
the parameter estimates for $\Omega_{m0}$, $H_0$, and $r_{ drag}$ obtained from 
our model are very close (within $2\sigma$) to the corresponding parameters for 
the case of $\Lambda$CMD model across all data combinations considered. It is 
tempting to conclude that the corresponding Hubble constant value gets reduced 
in comparison with the $\Lambda$CDM for all  dataset combinations that include 
the Pantheon+/SHOES dataset, however this reduction is well within $1\sigma$ 
levels, thus it remains statistically insignificant. Also, the known 
correlation between $r_d$ and $H_0$ appears, i.e. increased $H_0$ corresponds 
to smaller $r_d$ values, which is a manifestation of the so-called 
multidimensionality of the Hubble tension, see e.g \cite{Pedrotti:2024kpn}.

Hence, although the WGB framework allows small shifts
in the inferred values of $H_0$ and $\sigma_8$, the
present late-Universe analysis does not provide
statistically significant evidence for an alleviation of the
corresponding cosmological tensions. In particular, the
deviations from the concordance $\Lambda$CDM
parameter values remain well within the 1$\sigma$
posterior uncertainties for all dataset combinations
considered here.

A more robust assessment of the possible impact of the
WGB mechanism on the $H_0$ and $\sigma_8$ tensions
would require a full analysis including CMB data and
early-Universe observables, which lies beyond the
scope of the present work.

We note that the parameter $C_n$ effectively encodes the
combined contribution of several astrophysical quantities
related to black-hole formation and merger histories.
Hence, the posterior constraints obtained in the present
analysis should be interpreted as effective constraints on
the overall WGB mechanism, rather than as direct bounds
on the individual astrophysical parameters entering
Eq. (\ref{Cdef}).

Regarding the model comparison (see Tab. \ref{tab:info_criteria_wgb_lcdm}), the 
general picture is that the concordance model is consistently preferred by the 
data, in all subsets considered. Regarding the WGB scenario, the Model I is 
generally preferred over Model II, where the latter is in moderate tension with 
the data.  For the cases of CC/Pantheon+/SH0ES/BAOs and Pantheon+/SH0ES/BAOs, 
BIC criterion points to strong evidence against Model II, while AIC and DIC 
criteria show moderate tension in all cases. This situation is common in 
cosmological model selection, i.e \cite{Anagnostopoulos:2018jdq}, as the BIC 
criterion penalizes heavily on the extra free parameters. In contrast, for Model 
I we observe $\Delta IC \lesssim 2$ in all cases, which corresponds to 
statistical compatibility with the concordance model.

We use the aforementioned data and the expressions \eqref{model 1st Friedmann}, 
\eqref{SFR w z} to reconstruct the $H(z)$ and $w_{DE}$ respectively. In 
Fig.\ref{fig:hubblerec} we present the reconstructed evolution of the Hubble 
parameter for both models of the WGB scenario (red line), obtained directly by 
re-sampling the posterior parameter distribution in our analysis. The grey 
shaded areas, correspond to $1\sigma$ (deep
grey) and $2\sigma$ (light grey) regions. The black dashed line correspond to 
the Hubble parameter of the $\Lambda$CDM scenario. In the context of late 
Universe, both models are compatible within 1$\sigma$ with the concordance one, 
while they are able to reproduce the observed expansion history. with values 
that are slightly higher than those of the $\Lambda$CDM model.

In Fig. \ref{fig:w_DE}, we plot the evolution of the DE equation-of-state 
parameter $w_{DE}$ within the WGB framework for Model I (red line) and for Model 
II (blue line) derived from the best-fit parameters from all datasets 
considered. Note that the equation of state parameter for Model I shows phantom 
behavior, stabilized to $w_{DE} = -2$ for $z \sim 4$ and onward. In contrast, 
Model II allows for both phantom and quintessence behavior, the former being 
preferred by CC/Pantheon+/SH0ES/BAOs and Pantheon+/SH0ES/BAOs datasets and the 
latter by CC/BAOs dataset. In all cases, for $z \sim 8$, the extra contribution 
in the DE component from WGB mechanism becomes negligible and the model goes to 
$\Lambda$CDM.

We have solved numerically the coupled system of scalar potential, matter and 
DE perturbations \eqref{eq:pert_phi}, \eqref{eq:overd_eq_m_z} and 
\eqref{eq:delta_ddelta_DE}, employing the best fit parameters, as they are 
presented in Tab.~\ref{params_wgb_lcdm}, and for various values of the DE 
acoustic speed $c_e \in [0,1]$. As a general comment, the impact of each best 
fit value from Tab.~\ref{params_wgb_lcdm} is less than an order of magnitude of 
the observational 1$\sigma$ errors, for both Models I, II. In 
Fig.~\ref{fig:fs8_c_e} we have plotted the $f\sigma_{8}$ observable for 
$\Lambda$CDM and the WGB model for the three parameter sets, on top of the 
latest growth data \cite{Avila:2022xad}. We can see that the WGB model remains 
very close to the concordance model, well inside the observational bounds. In 
particular, Model II coincides with the $\Lambda$CDM, while Model I provides 
larger values for the nominal $\sigma_{8,0} = 0.81$ value employed at the plot. 
  This is an interesting feature, as it suggests that
Model I may allow observationally compatible fits
with slightly smaller values of $\sigma_{8,0}$.  Further analysis presented 
in Fig.~\ref{fig:Delta_fs8} shows that the effect of the $c_e$ value on the 
differences $\Delta f \sigma_8 =  f \sigma_{8}(c_{e}=1) - f 
\sigma_{8}(c_{e}=0)$ is relatively small, i.e. $\Delta f \sigma_8 / f 
\sigma_{8}\sim 10^{-6}$, for all the three datasets. Moreover, the dependency of 
the $f\sigma_{8}$ on the $c_e$ is more than an order of magnitude less than the 
typical $1\sigma$ error of the available data points \cite{Kazantzidis:2018rnb}. 
The latter is expected as the equation-of-state parameter, $w_{DE}$ is 
consistently different than $w=0$ for all z, thus the DE clustering must be 
negligible.

Similarly, in Fig.~\ref{fig:Geff_c_e} we have plotted the quantity 
$(G_{eff}/G_{N}-1)$ per redshift, for both models, utilizing the best fit 
parameters of the most complete dataset considered, i.e. 
CC/Pantheon+/SH0ES/BAOs. The results lie well within the observational bounds 
reported in \cite{Lamine:2024xno}, where it was found that $G_{\text{eff}}/G_N - 
1 = 10^{-2} \pm 2 \cdot 10^{-1}$. We further analyzed the dependence of 
$G_{\text{eff}}/G_N - 1$ on the DE acoustic speed $c_e$ for the three dataset 
parameters, as presented in Fig.~\ref{fig:Geff_c_e}. For both models, the 
absolute differences remain within three orders of magnitude of the 
aforementioned constraint. As expected, the deviation increases as clustering 
becomes stronger. The strong suppression of the $c_e$ effect is a direct 
consequence of the fact that DE clustering in both WGB models is very weak. 
Another distinction between Model I and Model II is that, while Model II 
exhibits an increment in $G_{\text{eff}}$ (i.e., positive $G_{\text{eff}}/G_N - 
1$), Model I exhibits a decrement. Assuming a physical origin for the $\sigma_8$ 
tension, this behavior could, in principle, contribute to its alleviation 
\cite{Kazantzidis:2018rnb}.

Despite the fact that the WGB cosmology remains close
to $\Lambda$CDM at the background and perturbation
levels, the framework possesses several distinctive
features that could potentially allow for observational
discrimination in the future. In particular, unlike the
standard cosmological constant scenario, the effective
dark-energy sector is directly connected to black-hole
formation and merger histories through the
gravity-thermodynamics conjecture and horizon topology.
Consequently, the late-time evolution of dark energy is
not fundamentally constant, but instead depends
explicitly on the cosmic star-formation history through
the function $\psi(z)$.

An additional distinctive feature of the framework is
that phantom behavior may emerge effectively without
introducing exotic scalar fields or negative kinetic
terms. Moreover, depending on the sign and magnitude
of the effective parameter $C_n$, the WGB mechanism
may either increase or reduce the effective cosmological
constant contribution, allowing in particular the
possibility of realizing negative effective cosmological
constants.

Finally, although current observational uncertainties
render the WGB predictions nearly degenerate with
$\Lambda$CDM at the perturbation level, future
improvements in black-hole population measurements,
merger-rate observations and late-time structure-growth
data could potentially provide additional observational
discriminants for the model.

In summary, the fact that $G_{eff}(z) \simeq G_N$ and the DE clustering is 
negligible, shows that WGB mechanism cannot describe Dark Matter - like 
effects. The latter holds for both astrophysical and cosmological scales.  
This is another important difference from the
Topological DE model of \cite{Anagnostopoulos:2025tax}, which can exhibit
transitions between Dark Energy and Dark Matter
and may effectively account for part of the dark
matter sector. Although, in the case of 
small mass BHs (e.g primordial BHs), the WGB mechanism could possibly produce 
more rich phenomenology. However the latter endeavor is left for a future 
project.

This behavior highlights an important phenomenological
difference between the WGB framework and other
topological dark-energy scenarios. In particular,
while certain realizations of Topological Dark Energy
(TDE) models may effectively interpolate between
dark-energy-like and dark-matter-like behavior through
stronger clustering properties, the WGB mechanism
remains much closer to a smooth dark-energy sector.
This difference originates from the distinct physical
interpretation underlying the two approaches: in WGB
cosmology the effective dark-energy contribution is
associated with black-hole horizon formation and
mergers through the gravity-thermodynamics
conjecture, whereas in TDE cosmology it emerges from
spacetime-foam-related topological effects.

\section{\label{Conclusion}Conclusions}
In this work, we explored the observational implications of Wald-Gauss-Bonnet 
(WGB) topological dark energy. This modified cosmological framework arises from 
applying the gravity-thermodynamics conjecture to the apparent horizon of the 
Universe, where the Wald-Gauss-Bonnet entropy replaces the standard 
Bekenstein-Hawking one. Assuming that the apparent horizon is topologically 
connected to interior black hole (BH) horizons, we derived modified Friedmann 
equations that depend on BH formation and merging rates, which is approximated 
to be proportional to the star formation rate. Consequently, the modified 
Friedmann equations depend on the Gauss-Bonnet coupling constant 
$\tilde{\alpha}$ and on known astrophysical parameters.
In the general case with $\Lambda \neq 0$ (Model II), the WGB mechanism 
introduces an additional contribution to the cosmological constant, which can 
be either positive or negative. In the particular case of a negative 
contribution, we show that negative cosmological constant can be realized. In 
the case where $\Lambda = 0$ (Model I), the WGB mechanism solves the 
cosmological constant and coincidence problems by its construction,  while it 
does not possess an explicit $\Lambda$CDM limit.

We carried out a Bayesian likelihood analysis using background data from 
SNIa/SH0ES, BAO, and cosmic chronometers (CC) to obtain the posterior 
distribution of the free parameters of WGB cosmology.  Following standard model 
selection criteria, we found that while both models are less favored than the 
concordance $\Lambda$CDM model, Model I ($\Lambda = 0$), remains statistically 
compatible with the latter, 
while  Model II ($\Lambda \neq 0$) is in moderate tension. Nevertheless, 
the novelty of the WGB framework does
not primarily arise from large deviations from the
concordance model, but from the fact that the effective
dark-energy sector emerges dynamically from
black-hole formation and merger histories through
topological and thermodynamical considerations.

We developed perturbation analysis in the context of effective fluid approach 
and investigated the growth of structures and the evolution of matter 
overdensities. For both WGB models considered here, the effective Newton’s 
constant marginally differs from the standard value ($10^{-3}$) and DE 
clustering is negligible. From the latter, we conclude that the WGB mechanism 
cannot mimic Dark-Matter-like effects in the cosmological realm, thus we still 
need an extra physical mechanism to describe Dark Matter. In this respect, 
the phenomenology of the WGB framework
differs substantially from certain Topological Dark Energy
scenarios in which topology-induced effects may also
generate effective dark-matter-like behavior through
enhanced clustering properties.  
Motivated by this, we consider it an interesting
direction for future work to apply the WGB
mechanism in the context of primordial black holes.

Regarding the thermal history of the Universe, Model II possesses both explicit 
and asymptotic (large z) $\Lambda CDM$ limit, thus it maintains the standard 
thermal history. On the other hand, the effective Dark Energy component of Model 
I approaches zero at large redshifts, thus the radiation era is preserved.  
It is noteworthy that Model I shows phantom behavior,
a feature often associated with possible alleviation of
the Hubble and growth tensions in late-time cosmological
scenarios. Nevertheless, the present analysis indicates
that the deviations from the concordance $\Lambda$CDM
predictions remain statistically small, with the inferred
values of $H_0$ and $\sigma_8$ remaining compatible
within the corresponding 1$\sigma$ uncertainties.
In contrast, Model II shows quintessential behavior.
A full assessment regarding the possibility of tension
alleviation would require a dedicated analysis including
CMB and additional early-Universe probes, which is
left for future work.

\bibliography{biblio}

\end{document}